# Temporal Heterogeneities Increase the Prevalence of Epidemics on Evolving Networks

Luis E C Rocha and Vincent D Blondel

Department of Mathematical Engineering

Université catholique de Louvain, Louvain-la-Neuve, Belgium

Luis.Rocha@uclouvain.be

15 June 2012

**Abstract**

Empirical studies suggest that contact patterns follow heterogeneous inter-event times, meaning that intervals of high activity are followed by periods of inactivity. Combined with birth and death of individuals, these temporal constraints affect the spread of infections in a non-trivial way and are dependent on the particular contact dynamics. We propose a stochastic model to generate temporal networks where vertices make instantaneous contacts following heterogeneous inter-event times, and leave and enter the system at fixed rates. We study how these temporal properties affect the prevalence of an infection and estimate $\langle R_0 \rangle$, the number of secondary infections, by modeling simulated infections (SIR, SI and SIS) co-evolving with the network structure. We find that heterogeneous contact patterns cause earlier and larger epidemics on the SIR model in comparison to homogeneous scenarios. In case of SI and SIS, the epidemics is faster in the early stages (up to 90% of prevalence) followed by a slowdown in the asymptotic limit in case of heterogeneous patterns. In the presence of birth and death, heterogeneous patterns always cause higher prevalence in comparison to homogeneous scenarios with same average inter-event times. Our results suggest that $\langle R_0 \rangle$ may be underestimated if temporal heterogeneities are not taken into account in the modeling of epidemics.

**INTRODUCTION**

Living in society implies that individuals are constantly interacting with each other. Interactions may take different forms, but those involving proximity or direct contact are of special interest because they are potential bridges for propagating infections. Empirical data suggest that contact patterns and consequently the network structure depend on the environment and context of the contacts [1]. The way people find sexual partners [2, 3] is not necessarily the same as they interact in a conference [4], museum exposition [4], school [5], hospital [6,7], or simply in daily life [8]. In common, these interactions have the property of being highly heterogeneous in the sense that individuals interact to different people at irregular times [2–8]. In the jargon of networks, each person, for example, has a different number of contacts, belongs to a different network cluster, or interacts with others at different rates [1,9]. Most research on networks has assumed that contacts are fixed, or static, and the topology at different levels contains the relevant structures to regulate the spreading processes, like a viral infection [1,9]. This approach works well as a first approximation in many circumstances, including when the process propagates much faster than the network evolves [10] (e.g. contacts of school children or patients in hospitals within a day), or when the activity patterns are homogeneous. In fact, for some regular activities people seem to be predictable [11] and consequently tend to have the same partners and contact structure on a typical day. Nevertheless, in some contexts, for example, people involved in commercial sex [3], or in tourism business, or hospital staff and long-term hospitalized patients [6], are constantly contacting different individuals during short intervals. In other words, the contact patterns change faster than the infective state of a person. There is further evidence that contact patterns are not regular but follow heterogeneous inter-event times, which means that intense activity (sometimes called bursts) may be followed by longer intervals of isolation or inactivity [3, 4, 7, 12, 13]. This property makes it difficult to distinguish on empirical data whether an individual is absent for some period or simply left the system (i.e. censoring [14]).

There is extensive literature about temporal patterns being considered on mathematical models of propagation of infectious diseases, some examples include, dynamic contact patterns, with or without considering the distance between two contacts [8, 15–18], seasonal effects affecting Influenza [19, 20], pair formation models [17, 21], birth and death [19, 22]. The role of the heterogeneous inter-event times, on the other hand, has been addressed only recently by few theoretical and data-driven studies. In particular, it has been suggested that, in case of power-law inter-event time distributions, the number of new infected individuals in SI-dynamics decays as a power-law in the long-time limit irrespective of the degree distribution of the network [23, 24]. These studies are essentially concerned with the long-term effect of the spreading, and not with the early stages of an epidemic outbreak. A study using an empirical network of sexual contacts suggests that, in comparison to homogeneous contact patterns, heterogeneous contacts speedup the spread of simulated infections [25] while the opposite happens, though the relative difference is small, in a contact network of conference attendees [26]. These results on sexual contacts, conference attendees, and other similar studies using diverse communication patterns as proxies for contact networks and SI-like simulations [27–29], are based on contrasting the evolution of the infection dynamics in the original network with the same dynamics on randomized versions of the temporal network (null models) where the time stamps are reshuffled retaining some (e.g. daily patterns) or no temporal constraints. Therefore, the relative difference and importance of the temporal characteristics, to shape the spread of infections, depends on the randomization protocol and therefore, results may be misleading. One example is the fact that the average inter-event time changes after the randomization. If some individuals are in the system originally at times $\Delta t < t < T_{final} - \Delta t$ (with $0 < \Delta t < T_{final}$), the reshuffling of time stamps results on contacts being active at new times $0 < t^* < T_{final}$. This effect is observed in networks where newcomers are constantly entering and old members leaving the system [3, 6]. In that case, the randomization of the time stamps causes an increase in the average inter-event times because the vertices that originally appeared, for example, in the beginning of the network, now may be found at any time $t^*$. If samples of the empirical network are subsequently repeated to extend the original network in the time domain, this turnover mechanism creates artificial cyclic effects and possibly abrupt changes in the network since the identity of vertices in the final and initial parts of the network are substantially different.

To avoid conclusions based on particular samples of empirical networks, it is desirable to study the spread of infections in a minimalistic model containing only the temporal properties of interest. In this paper, we propose a simple and intuitive model of a temporal network, where the vertices are available ("on" state) only at certain times and inactive ("off" state) otherwise, co-evolving with an infection dynamics. We mainly focus in the susceptible-infective-recovered (SIR) epidemics but as relevant case-studies, we also present results of the susceptible-infective (SI) model, typically used in the previous studies on heterogeneous inter-event times, and a reinfection model given by the susceptible-infective-susceptible (SIS) dynamics.

**METHODS**

**TEMPORAL NETWORK MODEL**

A temporal network may be defined as a dynamic network where the vertices are available ("on" state) only at certain times and inactive ("off" state) otherwise [30]. The links are not fixed and change as time goes by. In our model, each vertex follows a stochastic process where subsequent "on" states depends on a certain inter-event time distribution $P(\Delta t)$. In other words, we have a process where the probability of a vertex being active at time $t$, depends on the last time $t'$ it was active, i.e. $\Delta t = t - t'$. For computational reasons, time is discrete $t = \{1, 2, ..., T_{final}\}$ and we generate the next "on" state when the vertex is active, i.e. if a vertex is active at $t'$, we select the next time it will be active $t$ sampling $\Delta t$ from $P(\Delta t)$ [31]. As soon as a vertex is active, it chooses uniformly between other active vertices, connects to one of them during one time step, and destroys the link afterwards turning back to "off" state. Such evolving network is therefore only constrained by the inter-event time and results in randomly mixed networks without degree heterogeneities and correlations; after few steps everyone have contacted everyone else at least once. The inter-event time distribution depends on the process of interest. For a realistic scenario, we use a power-law with exponential cutoff $P(\Delta t) \propto \Delta t^{-\alpha} \exp(-\beta_{het}\Delta t)$ ($\alpha >> \beta = 0.001$), hereafter referred to as $\tau_{het}$ case, which is a typical example of a class of broad distributions and is also claimed to describe empirical data of sexual contacts [3], close contacts [4, 7] and other forms of human communication [12, 13, 27–29]. This type of inter-event time distribution are suggested to appear e.g. due to nonhomogeneous Poissonian processes constrained by cyclic activity [12], or due to preferential queuing models where individuals choose to perform highest priority tasks more often than random tasks [13]. As a baseline for comparison, we consider the exponential distribution $P(\Delta t) \propto \exp(-\beta_{hom}\Delta t)$ with the same average $\langle \Delta t \rangle$ as $\tau_{het}$. Exponential inter-event times appear on Poissonian processes where the chance of being active depends only on the probability $\beta_{hom} = 1/\langle \Delta t \rangle$ (which means that $t$ and $t'$ are unrelated in the Poissonian case) that corresponds to homogeneous inter-event times ($\tau_{hom}$). Since our models use discrete time, we use the equivalent geometric distribution for the simulations. The most adequate functional form for the inter-event time distribution depends on the system of interest but for simplicity we use the two limiting scenarios of complete randomness (Poisson) and burstness (power-law).

Independently of the evolution of these contacts, we define a turnover probability to account for the removal of the vertex. For simplicity, the lifetime of a vertex also follows a Poissonian process, where the removal probability is $\beta_{death} = 1/\langle \Delta t \rangle_{death}$. The death time $t_{death} = t' + \Delta t_{death}$ of the new vertex is taken from the distribution $P(\Delta t_{death}) \propto \exp(-\beta_{death} \Delta t_{death})$. A new vertex automatically replaces (birth) a removed vertex (death) in order to maintain the total number of vertices $N$ constant. Therefore, the prevalence is given by $\Omega(t) = \sum_{i=1}^{N} I(t)/N$. As initial conditions, all vertices start with random values for $t$ and $t_{death}$. The initial transient quickly disappears after few time steps and is discarded.

**EPIDEMICS MODELS**

On top of the evolving network, we define an infectious process. We focus our study on general properties of the susceptible-infected-recovered (SIR) dynamics (adequate to model e.g. one wave of Influenza, Measles, or Hepatitis B) [19, 21]. For comparison to previous studies based on empirical networks, we consider the limit case of susceptible-infected (SI) dynamics (which is also valid as a first approximation of an early stage of HIV propagation). Due to its importance to model several common infections (e.g. Influenza, Chlamydia, Gonorrhea) by accounting for the possibility of reinfection, we also investigate some properties of the susceptible-infective-susceptible (SIS) model [19, 21]. The study of specific infections with realistic parameters and the most proper compartmental models go beyond the scope of the present article, therefore, we focus on more theoretical aspects of these standard models to understand the impact of heterogeneous temporal contacts on general spread dynamics.

On the SIR model, a vertex infects instantaneously a susceptible partner, upon contact, with probability $\lambda = 1$ ($\lambda$ is the per-contact infection probability), and an infected vertex is completely recovered after $\Delta t_I$ time steps. In the SI model, $\Delta t_I \to \infty$, and in the SIS model, the vertex becomes susceptible after the infective period $\Delta t_I$ allowing reinfections. Note that recover means that the vertex cannot infect or be infected, but still makes connections until being removed by death. Initially, one vertex is set to the infective state and the remaining $N-1$ vertices are set susceptible. Newcomers are always set to susceptible state. We generate 50 ensembles of the temporal network and for each ensemble, select all vertices active in the first time step; for example, $N = 1000$ gives on average $\sim 250$ active vertices that are used as infection seeds. This results on approximately $50 \cdot 250 = 12500$ different initial conditions.

**RESULTS**
**SIR DYNAMICS**
We initially study the evolution of the SIR dynamics on the temporal network and investigate the prevalence of the infection within the population according to the different temporal properties of the contact patterns and the parameters of the epidemics. In the absence of death ($\beta_{death} = 0$), the only temporal pattern is the heterogeneous inter-event times. Figure 1a shows the difference in the peak values $\Delta \Omega = \max \langle \Omega_{het}(t_{peak-het}) \rangle - \max \langle \Omega_{hom}(t_{peak-hom}) \rangle$ for the two scenarios of inter-event times. Varying the infective interval $\Delta t_I$ and the contacts heterogeneity, which is given by the slope $\alpha$ of the power-law distribution (see Supporting Information T1 for the relation between $\langle \Delta t \rangle$ and $\alpha$), raise two distinct regimes where the peak of prevalence is higher for heterogeneous contact patterns $\tau_{het}$ (positive values of $\Delta \Omega$) for small values of $\Delta t_I$ and $\alpha$, whereas higher values of the same parameters produce the opposite effect (negative values of $\Delta \Omega$), with higher prevalence for the homogeneous scenario $\tau_{hom}$. The difference in the peak prevalence $\Delta \Omega$ reaches up to 54%. The irregular contact patterns also cause an earlier peak of the maximum prevalence (Fig. 1b, where $\Delta t_{peak} = t_{peak-het} - t_{peak-hom}$ such that negative values mean that the peak of $\tau_{het}$ occurs earlier than of $\tau_{hom}$). For $2 < \alpha < 3$, which covers values observed on empirical systems [3, 4, 7], $t_{peak}$ occurs 20 to 40 time steps earlier for moderate $\Delta t_{death}$. The region around $\alpha < 2$ in Figure 3b essentially comprises cases where the infection dies out quickly ($t_{peak-hom} \approx 0$) for homogeneous patterns but not for heterogeneous.

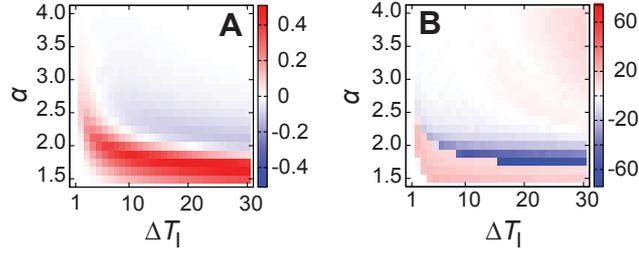

**Figure 1. Value and time of the maximum prevalence for SIR.** Difference in the (a) value of maximum prevalence $\Delta\Omega$ $\Delta\Omega$, and in the (b) time of the maximum prevalence $\Delta t_{peak}$ $\Delta t_{peak}$, for SIR dynamics with various infective intervals in the absence of death.

The evolution of the SIR epidemics in the two regions identified in Figure 1 is shown in more details using two specific configurations with $\alpha = 2.5$, $\Delta t_I = 5 \approx 3\langle\Delta t\rangle$ (Fig. 2a) and $\Delta t_I = 10 \approx 6\langle\Delta t\rangle$ (Fig. 2b). The infection growth has a similar shape for both inter-event times with a slightly broader spread for $\tau_{het}$. The final fraction of susceptible vertices is higher for $\tau_{het}$ irrespective of the values of $\Delta t_I$, which suggests that the earlier peak is responsible for avoiding the infection of the entire network. The introduction of birth/death reduces the peak prevalence on both contact patterns but has more impact on the homogeneous case (Figs. 2c-f). For moderate turnover rate $\beta_{death}$, i.e. $\Delta t_{death} \approx 6\langle\Delta t\rangle$ (Figs. 2c and 2d), susceptible newcomers are responsible to create a small second wave of epidemics (peak is ~40% smaller than the first wave for both configurations), followed by a small oscillation of low prevalence for both cases but still higher for heterogeneous contacts. Similar results are observed for other configurations with $\Delta t_{death} \approx 6\langle\Delta t\rangle$ and $\alpha > 2$ (See Supporting Information T1), but $\Delta t_{peak}$ decreases for increasing $\alpha$. When the turnover is higher $\Delta t_{death} \approx 3\langle\Delta t\rangle$ (Figs. 2e and 2f), the prevalence grows monotonically to a state of constant prevalence and high fraction of susceptible individuals (60% to 80% of vertices for the configurations shown).

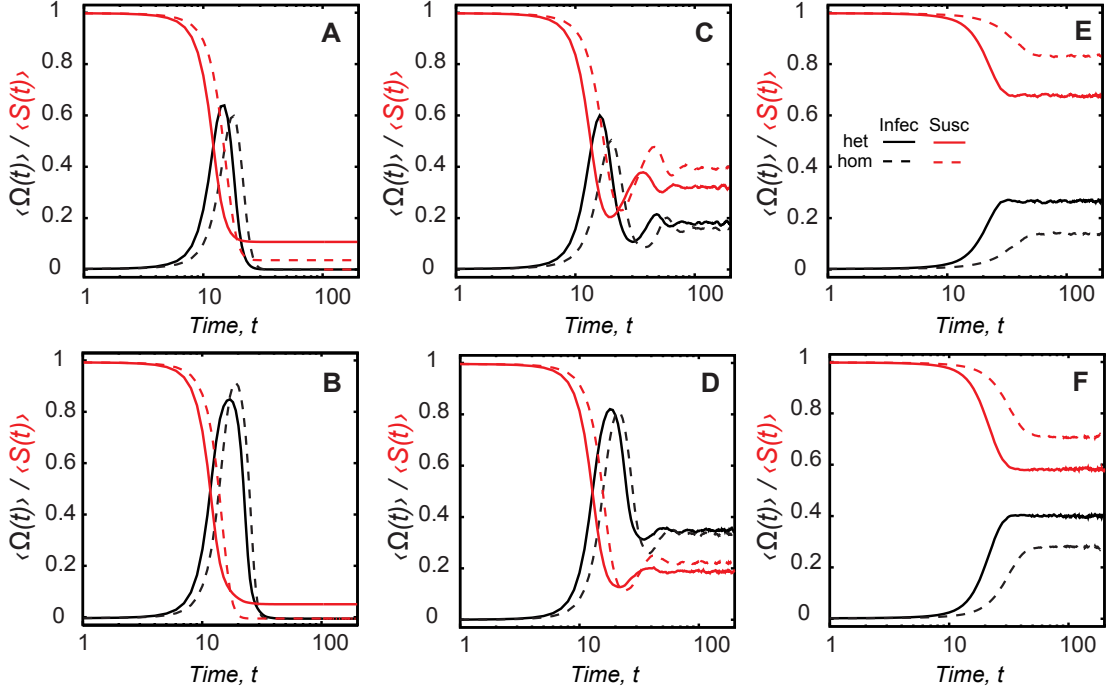

**Figure 2. SIR epidemics on temporal networks following heterogeneous ($\alpha = 2.5$) and homogeneous inter-event time distributions.** Curves correspond to the fraction of infected (i.e. the prevalence – blue lines) and fraction of susceptible individuals (red lines). First row corresponds to the configuration with $\Delta t_I = 5$ and second row to the configuration with $\Delta t_I = 10$ $\Delta t_I = 10$. First column shows the absence of death, second and third columns the presence of death, respectively, $\Delta t_{death} = 5$ $\Delta t_{death} = 5$ and $\Delta t_{death} = 10$ $\Delta t_{death} = 10$. The x-axis is in log-scale.

The introduction of the two temporal aspects in the dynamic network creates unique infection pathways within the network even though every vertex contacts every other vertex within a short time interval. These infection routes create a diversity of possible outbreaks, enhancing the dependence of the outbreak with the initial conditions. Both in the absence and in presence of death, the distribution of outbreaks $P(\Omega)$ is broader for the $\tau_{het}$ in comparison to $\tau_{hom}$, but still with characteristic values (Fig. 3). The diversity increases with the death rate (Fig. 3b). Furthermore, in case of death, few vertices are removed before infecting their contacts, which results in a number of null outbreaks (Fig. 3b). In both scenarios, i.e. with and without death, there is a considerable overlap of the distributions indicating that according to the initial conditions and network ensembles, outbreaks might result in $\Omega_{hom} > \Omega_{het}$, which is different than the characteristic behavior of $\langle\Omega_{het}\rangle > \langle\Omega_{hom}\rangle$.

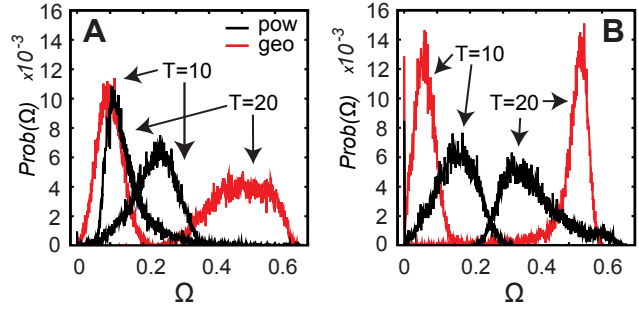

**Figure 3. Probability of an outbreak size for a random initial infection seed.** Fraction of times $P(\Omega)$ an outbreak $\Omega$ is observed at two different times $t = 10$ and $t = 20$. The results correspond to SIR model with $\Delta t_I = 5$ and $\alpha = 2.5$, in (a) the absence, and (b) presence of death $\Delta t_{death} = 20$.

**SI AND SIS DYNAMICS**

As two specific examples of infection dynamics, we study the impact of the temporal effects using the SI and SIS epidemics models. For the SI dynamics (Fig. 4a), in case of absence of death, the heterogeneous patterns $\tau_{het}$ causes a higher prevalence during an initial period, where about 85% of the population is infected, followed by a slow increase till the remaining network is reached. The homogeneous scenario $\tau_{hom}$ causes a slower growth in the initial interval, however, allows the infection to reach the entire network much earlier. If death is included, $\tau_{het}$ causes higher prevalence for the entire period and also removes the asymptotic slow growth observed in the absence of death (Fig. 4a). The heterogeneous inter-event time maintains non-zero prevalence for a vast range of $\Delta t_{death}$, while homogeneous contact patterns may result on null average outbreaks, especially for $\Delta t_{death} \approx \Delta t$. The SIS model shows a similar result in the sense that an initial higher prevalence $\langle\Omega\rangle$ for $\tau_{het}$, in comparison to $\tau_{hom}$, is followed by a lower $\langle\Omega\rangle$ in the steady state after a short oscillatory transient (Fig. 4b). The removal of vertices has a stronger impact on the prevalence $\langle\Omega\rangle$ for the homogeneous case; for increasing turnover (i.e. decreasing $\Delta t_{death}$), after a transition point $\Delta t_{death}^*$, $\langle\Omega\rangle$ gets higher for $\tau_{het}$ in comparison to $\tau_{hom}$.

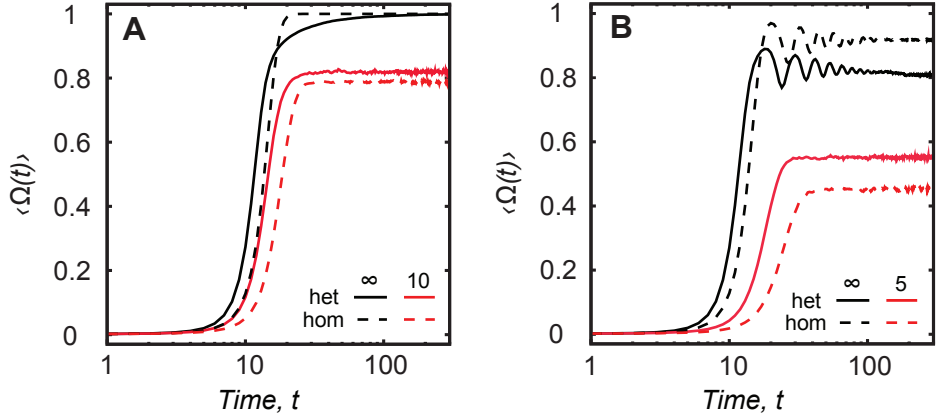

**Figure 4. Prevalence of the infection for SI and SIS epidemics.** The prevalence $\langle \Omega(t) \rangle$ in case of (a) SI and (b) SIS epidemics $\Delta t_I = 5$, considering absence (blue curves) and presence of death (red curves), for $\tau_{het}$ and $\tau_{hom}$ contact patterns. The x-axis is in log-scale.

## ESTIMATION OF $\langle R_0 \rangle$

We estimate $R_0$ by counting the number of secondary infections, during 200 time steps, produced by a single initially infected vertex in a completely susceptible population [19]. Figure 5 shows $\langle R_0 \rangle$ for heterogeneous and homogeneous contact patterns, in the absence and in presence of death. For SIR and SI (which is simply used as a limiting case of SIR when $\Delta t_I \to \infty$), the values of $\langle R_0 \rangle$ are higher in the presence of death. With the exception of SI in the absence of death (Fig. 5c), all other configurations show $\langle R_0 \rangle \approx 1$ for $\alpha \leq 2$ in the homogeneous case. For values of $\alpha \geq 3.5$, $\langle R_0 \rangle$ is indistinguishable for both contact patterns for all epidemics models, while in the range $1.5 \leq \alpha \leq 3$, $\langle R_0 \rangle$ is generally higher for $\tau_{het}$ in comparison to $\tau_{hom}$, with the exception of SI without death, where the opposite effect is observed.

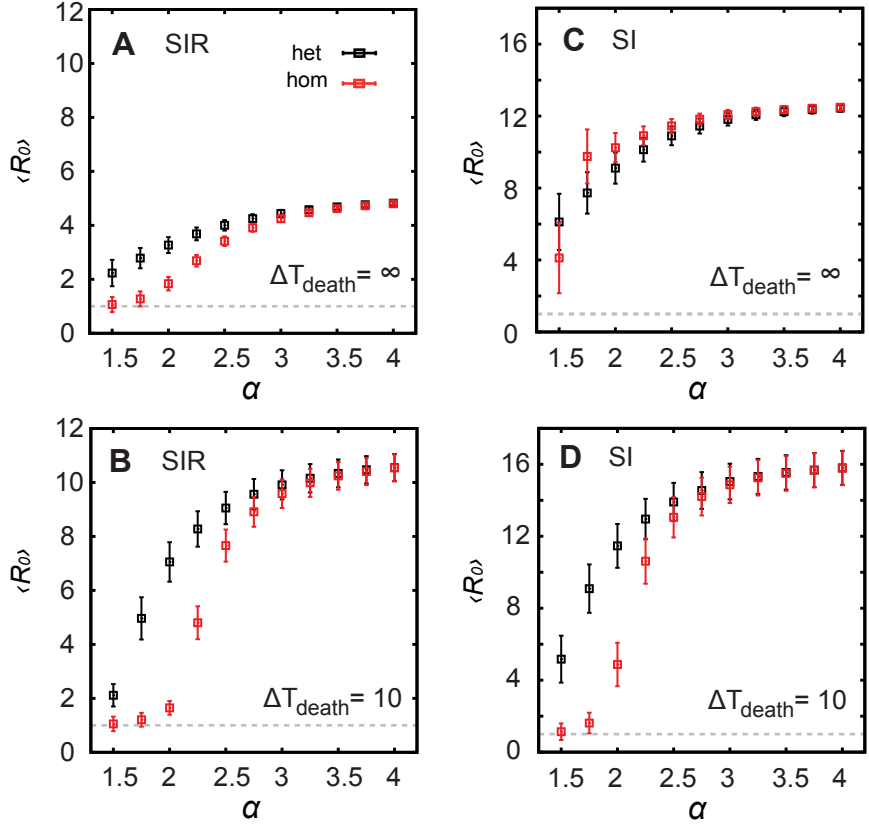

**Figure 5. Estimation of $\langle R_0 \rangle$ for SIR and SI epidemics models.** Numerical estimation of $\langle R_0 \rangle$ for SIR in the (a) absence and (b) presence of death; for SI in the (c) absence and (d) presence of death. The results for SIR are independent of the network size but SI depends on $N$ (see Supporting Information T1), therefore, we show results for $N = 8000$. We use $\Delta t_I = 5$ for SIR. Dotted lines correspond to $\langle R_0 \rangle = 1$. Bars represent the standard deviation.

**DISCUSSION**

Contact patterns are characterized by a high degree of heterogeneity both at topological and temporal levels [1–13]. An individual typically interacts with a number of partners during a period of time, but the interval between two interactions does not necessarily follows a characteristic value [3–6]. Furthermore, given the same period of time, some individuals may enter or leave the system at different times [3, 19, 22]. The two observations raise the question about to which extend modeling high-resolution temporal information affects the dynamics of spread of infections. As much as the contact structure (e.g. degree distribution [32] or community structure [33]) shapes the spreading of infections, heterogeneous inter-contact times also influence the dynamics of infection propagation [25, 26]. These questions have not been much explored though, and indeed, there are some contrasting conclusions about the actual impact of heterogeneous contact patterns on propagation of simulated infections on empirical network structures.

To contribute on the understanding of the role of temporal network structure on the propagation of infections, we introduce a simple and intuitive model of temporal network where the vertex dynamics is only constrained by the inter-event time distribution and by an independent probability of vertex removal. Within this simplified framework, the contribution of the temporal dynamics on the spread of infections becomes evident, since there are no competing structures, as for instance, degree heterogeneity, community structure, or assortativity [1, 9, 32, 33].

Our results show that for the 3 epidemics models considered (SIR, SI, SIS), the prevalence curve can be divided into 2 regimes. The first part is characterized by a faster, steeper growth of the fraction of infected vertices in case of heterogeneous contact patterns. After this initial growth, a second regime is identified whose characteristics depend on the death rate and on the epidemics model. In the absence of death, the prevalence of the infection is higher for homogeneous contact patterns. This result occurs because in a completely susceptible population, an infected vertex quickly, contacts several other vertices due to bursts and consequently the epidemics take off earlier. However, as time goes by, the longer inactive intervals due to the broad distribution of inter-event times results that the probability to find a susceptible vertex decreases and the heterogeneous contact patterns slowdown the spread of the infection. In our model, therefore, an initial speed up is followed by a slowdown in the asymptotic limit, which is in agreement with previous empirical and theoretical results [23–25, 29]. It may be that other broad but flatter distributions, not strictly power-laws [26–28], led to slower growth in the earlier stages of the epidemics. One should also expect that a high level of clustering combined with temporal structures may also slowdown the spread [29].

When death is included, infected vertices are replaced by susceptible vertices. Increasing the death rate causes a higher replacement of infected vertices, leaving more susceptible vertices available for infection. Consequently, on average it is expected that the same vertex, following heterogeneous inter-event times, contacts more different vertices during its lifetime than if following homogeneous inter-event times, even though the average inter-event time is the same for both scenarios, as defined in the model. Therefore, the difference in the prevalence of infection, by considering both contact patterns, is more pronounced in networks with high vertex turnover rate. This is the case of sexual networks related to commercial sex [3] or contact networks of people hospitalized in the same ward [6] where the percentage of active vertices detected both in the initial and final 5% of the sampled edges is, respectively, 0.3% and 0.8% [34]. On the other hand, closed communities [4, 34] or for example communication networks may have low turnover rate [34]. For SIR, the heterogeneous contacts act as a natural way of avoiding infection of the entire network. Due to newcomers, the fraction of susceptible vertices remains constant after the first wave of infection, but reaches larger values for increasing vertex turnover. Temporal structures therefore may be used to exploit new vaccination protocols based on behavioral characteristics of the population [34–36].

We have also estimated $\langle R_0 \rangle$, the average number of secondary infections produced by a single infective source in a completely susceptible population. For different levels of heterogeneity (measured by the exponent α of the power-law), we generally found $\langle R_0 \rangle_{SIR} < \langle R_0 \rangle_{SI}$. The difference in $\langle R_0 \rangle$ for heterogeneous and homogeneous contact patterns is larger for $\alpha \leq 2.5$. In the absence of death, for all values of $\alpha > 1.5$ in the SI dynamics, $\langle R_0 \rangle$ is larger for homogeneous contacts. In all other configurations, heterogeneous patterns result on higher values of $\langle R_0 \rangle$. In the range of parameters meaningful for real contact patterns, i.e. $2 < \alpha < 3$, our model gives $\langle R_0 \rangle$ values close to those estimated for common diseases [37, 38]. Nevertheless, the important result is that according to our results, traditional models that assume temporally homogeneous contacts and fixed population generally underestimate the value of $\langle R_0 \rangle$. This effect is particularly relevant in systems where the dynamics of contacts are highly heterogeneous, as is the case of people involved in commercial sex, hospitals, or tourism related activities.

We have proposed a simple temporal network model with heterogeneous inter-contact times and birth/death dynamics. The simulation of standard epidemics models (SIR, SI, SIS) has shown that generally, these irregular contacts are responsible to earlier and higher outbreaks in comparison to homogeneous temporal activity, although slowing down may be observed under some limiting configurations as the asymptotic limit in the SI–dynamics. Our results further suggest that neglecting the temporal heterogeneity underestimates the values of $\langle R_0 \rangle$. In conclusion, irregular temporal patterns affect significantly the spread of infections and further research is needed to understand the combined role of topological and temporal correlations on the emergence of epidemics.

**ACKNOWLEDGEMENT**

LECR is beneficiary of a FSR incoming post-doctoral fellowship of the Academie universitaire Louvain, co-funded by the Marie Curie Actions of the European Comission. Computational resources have been provided by the supercomputing facilities of the Université catholique de Louvain (CISM/UCL) and the Consortium des Équipements de Calcul Intensif en Fédération Wallonie Bruxelles (CECI) funded by FRS-FNRS.

**Supporting Information for:**

**Temporal Heterogeneities Increase the Prevalence of Epidemics**

**on Evolving Networks**

Luis E C Rocha and Vincent D Blondel

Department of Mathematical Engineering

Université catholique de Louvain, Louvain-la-Neuve, Belgium

Luis.Rocha@uclouvain.be

15 June 2012

## 1. Values of $\langle\Delta t\rangle$ used in the simulations

In the simulations, we use the same $\langle\Delta t\rangle$ for the heterogeneous and homogeneous contact patterns. For a given $\alpha$, we calculate $\langle\Delta t\rangle$ (Fig. S1) by sampling 100000000 (one hundred million) values from $P(\Delta t) \propto \Delta t^{-\alpha} exp(-\beta_{het}\Delta t)$ (where $\beta_{het}$ is fixed and $\alpha \gg \beta_{het} = 0.001$). This value of $\langle\Delta t\rangle$ is used in the exponential distribution $P(\Delta t) \propto exp(-\beta_{hom}\Delta t)$ by setting $\beta_{hom} = 1/\langle\Delta t\rangle$. In practice, since we use discrete times, times are sampled from a geometric distribution. Random generators can be found on numerical packages for diverse programming languages, otherwise, see ref. [R1].

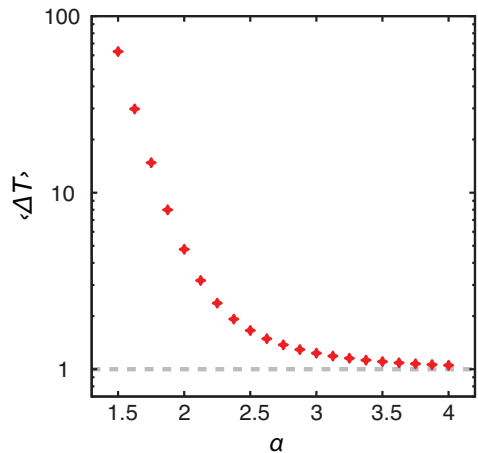

**Figure S1:** Values of $\langle\Delta t\rangle$ for a given $\alpha$ as obtained by sampling 100000000 values from the power-law (with cutoff) distribution. y-axis is in log-scale.

## 2. Effect of varying $\Delta t_{death}$ on the SIR dynamics

In Figure S2, we fix the infective stage at $\Delta t_I = 5$ and vary the death rate $\Delta t_{death}$ for 3 different values of α, to show the difference in the prevalence for the two scenarios of inter-event times during the initial 100 time steps of the epidemics. These results are complementary to Fig. 2 in the main text. For all values of α in the interval $1.5 \leq \alpha \leq 4$, the pattern of prevalence is similar, with an initial peak for the case of $\tau_{het}$, followed by a later peak in case of $\tau_{hom}$. For α = 2 and for small lifetimes, the heterogeneous case $\tau_{het}$ is always higher because it can sustain the epidemics while the case of homogeneous $\tau_{het}$ contact patterns results on multiple null outbreaks. For α = 2.5 and larger values of $\Delta t_{death}$, the results are similar to the case of absence of deaths, i.e. there is an initial interval where $\tau_{het}$ prevails causing higher outbreaks, followed by a stage where $\tau_{hom}$ gives higher outbreaks. On the other hand, for decreasing $\Delta t_{death}$, the sign of $\Delta\Omega$ alternates between the two scenarios $\tau_{hom}$ and $\tau_{het}$ during the initial 100 time steps. The second interval of higher prevalence for $\tau_{hom}$ (the second wave) appears earlier with decreasing $\Delta t_{death}$. This behavior is similar for values of α > 2.5 but the absolute values of $\Delta\Omega$ are smaller. On the other hand, for α = 2.25, the prevalence of the homogeneous case is higher only during one interval with $\tau_{het}$ giving higher outbreaks most of the time.

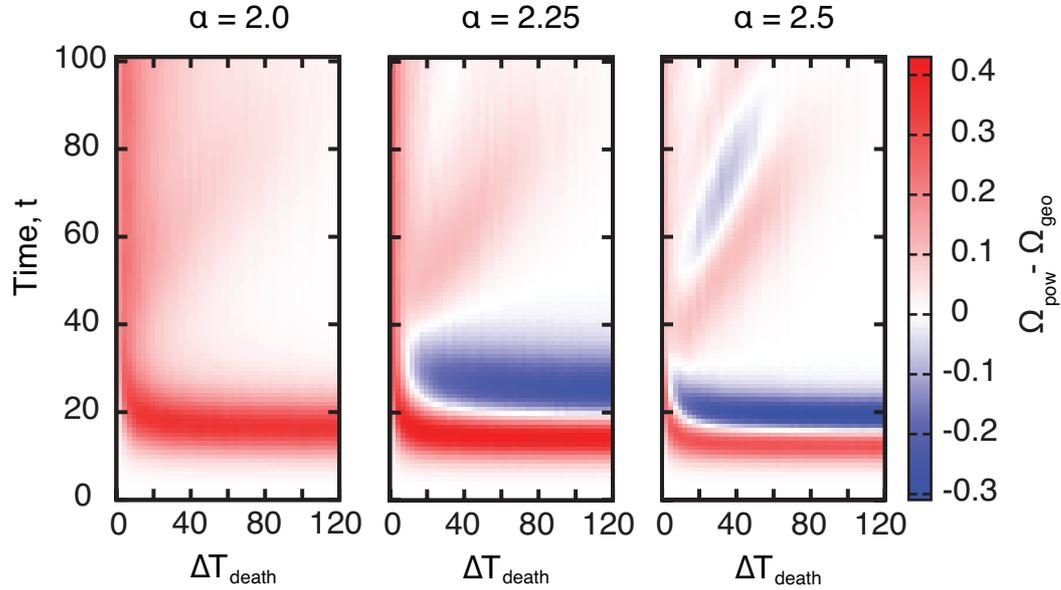

**Figure S2:** The difference in the prevalence $\Delta\Omega$ for the heterogeneous and homogeneous contact patterns for different values of α. x-axis corresponds to the lifetime $\Delta t_{death}$ and the y-axis corresponds to the time steps.

### 3. Finite-size analysis of $\langle R \rangle_0$ for SIR and SI dynamics

We perform the simulations of the epidemics on the temporal network for different sizes $N$ of the network to see if there is a dependence of $\langle R \rangle_0$ and $N$. By using values of $N = \{1000, 2000, 4000, 8000, 16000\}$, we see that our estimation of $\langle R \rangle_0$ is independent of the network size for SIR epidemics, and therefore, a reliable estimate (Fig. S3). For SI epidemics, $\langle R \rangle_0$ increases asymptotically with the network size, which is expected since once infected, a vertex will continue infecting throughout the dynamics. Strictly speaking $\langle R \rangle_0$ is inadequate to characterize the SI dynamics long after the initial infection, but is included as a comparison means between the two contact patterns (Fig. S4).

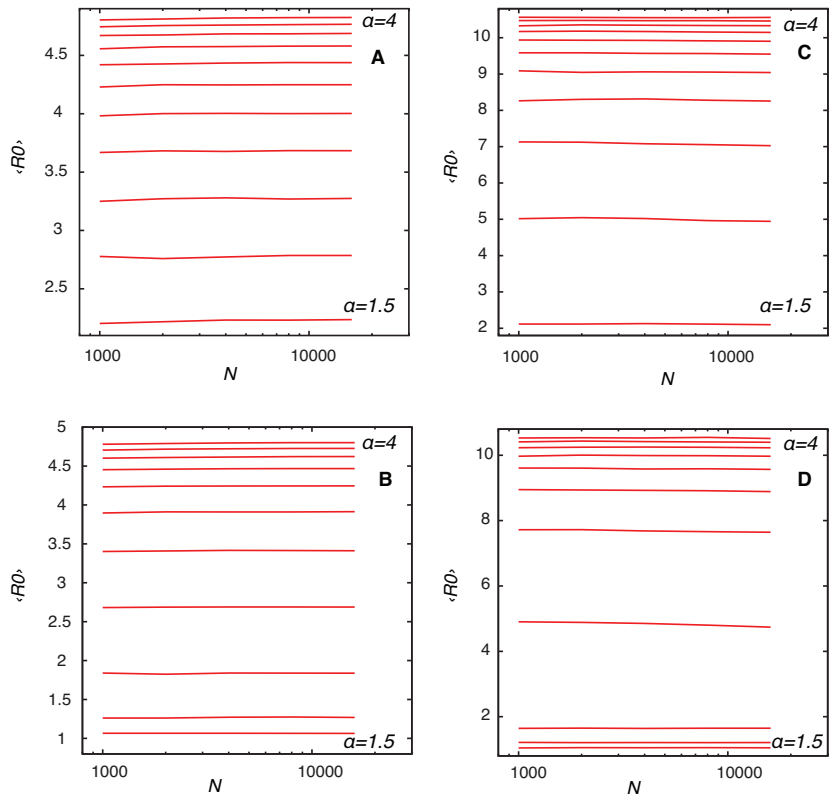

**Figure S3:** Estimation of $\langle R \rangle_0$ for SIR epidemics for different sizes $N$ of the temporal network. (a) SIR on heterogeneous network in the absence of death; (b) SIR on homogeneous network in the absence of death; (c) SIR on heterogeneous network in the presence of death with $\Delta t_I = 5$; (d) SIR on homogeneous network in the presence of death with $\Delta t_I = 5$. From bottom to top, each curve corresponds to a different growing value of $\alpha$.

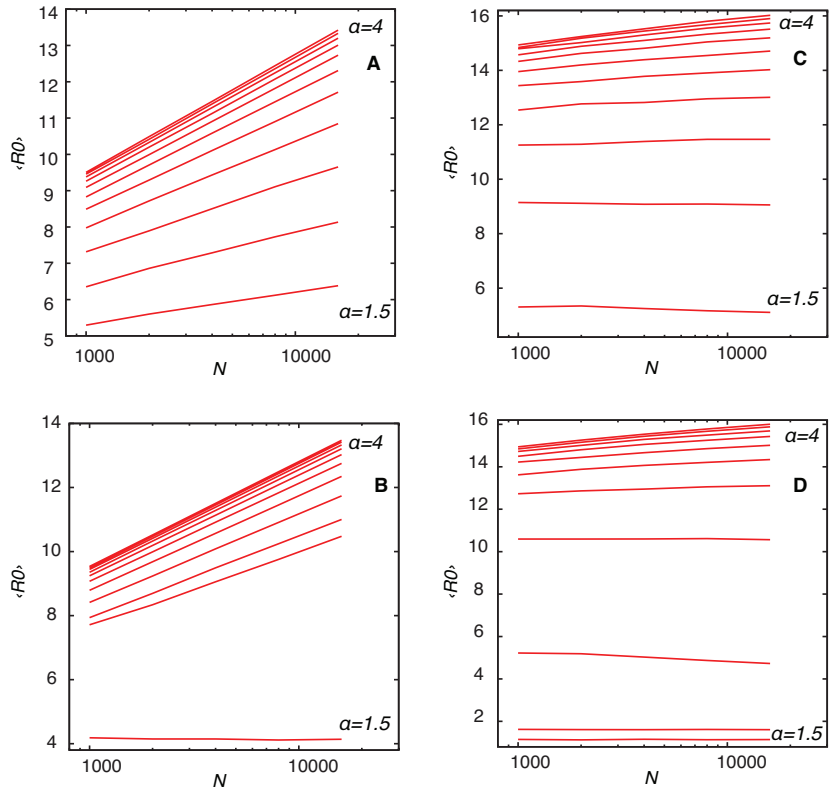

**Figure S4:** Estimation of $\langle R \rangle_0$ for SI epidemics for different sizes *N* of the temporal network. (a) SI on heterogeneous network in the absence of death; (b) SI on homogeneous network in the absence of death; (c) SI on heterogeneous network in the presence of death with $\Delta t_I = 5$; (d) SI on homogeneous network in the presence of death with $\Delta t_I = 5$. From bottom to top, each curve corresponds to a different growing value of $\alpha$.